\documentclass[prepintt]{elsarticle}
\usepackage[mathcal]{euscript}
\usepackage[latin1]{inputenc}
\usepackage{latexsym}
\usepackage{amsmath}
\usepackage{hyperref}
\usepackage{amsmath}    
\usepackage{graphicx}   
\usepackage{verbatim}  
\usepackage{color}      
\usepackage{subfigure}  
\usepackage{hyperref}   
\usepackage{bm}
\usepackage{graphicx}
\usepackage{amssymb}
\usepackage{bbm}
\usepackage{float}
\usepackage{slashed}
\usepackage{amsfonts}
\usepackage{euscript}
\usepackage{amsmath}    
\usepackage{mathrsfs} 
\usepackage{bbding}
\usepackage{pifont}
\usepackage{cancel}

\begin{document}

\begin{frontmatter}

\title{Absence of cosmological constant problem in special relativistic field theory of gravity}

\author{Carlos Barcel\'o}
\address{Instituto de Astrof\'{\i}sica de Andaluc\'{\i}a (IAA-CSIC), Glorieta de la Astronom\'{\i}a, 18008 Granada, Spain}
\ead{carlos@iaa.es}

\author{Ra\'ul Carballo-Rubio}
\address{Instituto de Astrof\'{\i}sica de Andaluc\'{\i}a (IAA-CSIC), Glorieta de la Astronom\'{\i}a, 18008 Granada, Spain

Department of Mathematics \& Applied Mathematics, University of Cape Town, Private Bag, Rondebosch 7701, South Africa

SISSA, Via Bonomea 265, 34136 Trieste, Italy and INFN Sezione di Trieste}
\ead{raul.carballorubio@sissa.it}

\author{Luis J. Garay}
\address{Departamento de F\'{\i}sica Te\'orica, Universidad Complutense de Madrid, 28040 Madrid, Spain
\ead{luisj.garay@ucm.es}

Instituto de Estructura de la Materia (IEM-CSIC), Serrano 121, 28006 Madrid, Spain}

\begin{abstract}{The principles of quantum field theory in flat spacetime suggest that gravity is mediated by a massless particle with helicity $\pm2$, the so-called graviton. It is regarded as textbook knowledge that, when the self-coupling of a particle with these properties is considered, the long-wavelength structure of such a nonlinear theory is fixed to be that of general relativity. However, here we indicate that these arguments conceal an implicit assumption which is surreptitiously motivated by the very knowledge of general relativity. This is shown by providing a counterexample: we revisit a nonlinear theory of gravity which is not structurally equivalent to general relativity and that, in the non-interacting limit, describes a free helicity $\pm2$ graviton. We explicitly prove that this theory, known as Weyl-transverse gravity (unimodular gravity with explicit Weyl invariance), can be understood as the result of self-coupling in complete parallelism to the well-known case of general relativity. We discuss the absence of cosmological constant problem in this theory, highlighting that it provides a particular realization of previous arguments formulated in studies of the emergence of the gravitational interaction from condensed-matter-like models. Overall, we conclude that the consideration that gravity is mediated by a massless particle with helicity $\pm2$ does not inextricably lead to the cosmological constant problem.
}
\end{abstract}
\begin{keyword}
graviton; self-coupling; cosmological constant; vacuum energy; unimodular gravity; emergent gravity; bootstrapping
\end{keyword}

\end{frontmatter}

\tableofcontents
%

%------------------------------------------------
\section{Introduction}
%------------------------------------------------

The lack of gravitating properties of vacuum zero-point energies of matter can be considered as the only available glance to the realm of quantum gravity with which we have been experimentally rewarded up to now. However, the application of effective field theory arguments to the combination of the standard model of particle physics and general relativity strongly suggests that what is reasonable from the theoretical perspective is, indeed, the contrary \cite{Weinberg1989,Burgess2013,Polchinski2006}. The cosmological constant problem \cite{Weinberg1989,Burgess2013,Polchinski2006,Martin2012} is usually defined as the problem of finding a mechanism that forbids these energies to gravitate (or, more formally, prevents semiclassical radiative corrections to modify the value ultimately taken by the cosmological constant) while, at the same time, respects the low-energy physics in order to guarantee that no contradictions arise with the stringent experimental tests on deviations from general relativity.

It is generally expected that this problem can only be solved within a theoretical framework that combines the principles of general relativity and quantum mechanics. At the same time, any quantum gravity theory must reproduce the known classical laws of physics, and hence lead to general relativity in the long-wavelength limit \cite{Carlip2012}. This has led to several attempts to find mechanisms that can avoid this problem, while maintaining the structure of general relativity intact. It is fair to say that, after decades of research, there is no completely satisfactory model that satisfies these requirements; moreover, any of these attempts includes unknown physics \cite{Burgess2013}. The spirit of the present discussion is different. Here, we start from the well-established principles of quantum field theory in flat spacetime, namely the observation that gravity is mediated by gravitons, as well as the equivalence principle as reformulated by Weinberg \cite{Weinberg1964b}, and follow these principles to their ultimate consequences. The starting point for our discussion is then the special relativistic setting pioneered by Feynman and others in their investigations about the quantum properties of the gravitational interaction \cite{Feynman2002}.

In this work we point out that there is a loophole in the chain of arguments that assert that the only possible consistent theory of self-interacting gravitons (particles with helicity $\pm2$) is general relativity. This has nontrivial consequences: there exists an alternative nonlinear theory of gravity that arises due to the self-interactions of such a particle as well. To the best of our knowledge, it is the first time that this has been solidly argued. This theory is similar in spirit to unimodular gravity and it is denominated Weyl-transverse gravity in order to highlight its symmetries. The interpretation of its nonlinear character as the result of self-coupling makes this theory a legitimate candidate to describe the long-wavelength limit of a hypothetical quantum gravity theory. This position is strengthened by the fact that it does not present the cosmological constant problem \cite{Carballo-Rubio2015}: the (effective) cosmological constant is rendered stable against radiative corrections, while the remaining phenomenological aspects of the standard model and general relativity are respected.\footnote{Contrary to other proposals, based in some form of scale invariance, which imply constraints on the matter sector even classically. This is of central importance for our discussion.}

What makes this proposal attractive is that it involves only assumptions which are natural in quantum field theory. One could have reached it without previous knowledge of general relativity. In fact, being accustomed to the geometric description of gravity even hinders its obtention: as we point out, this prior knowledge caused previous approaches such as \cite{Deser1970,Boulware1975} to overlook this theory. The lesson to be drawn is that the issue of vacuum zero-point energies is no longer a problem if one takes the principles of special relativity and quantum mechanics seriously, considering that general relativity is just an effective theory valid at sufficiently low energies.

A few sentences may be useful in order to properly place this paper with respect to previous work in the subject. It represents a continuation of the research reported in \cite{Barcelo2014}, where it was discussed how unimodular gravity can be obtained as a solution to the self-coupling problem. To reach this result we had, however, to consider a constrained graviton field from the beginning of the construction. For this reason the reader could get the impression that it is really the imposition of this constraint what it is behind of the obtention of unimodular gravity, making this solution still marginal with respect of general relativity. The content of this paper shows that this impression is not correct and conveniently merges the results of \cite{Barcelo2014} and \cite{Alvarez2006,Blas2007}. We do not discuss here the legitimation of the entire construction from the perspective of special relativity, which is thoroughly covered in \cite{Barcelo2014}, including detailed comments on previous works.

%------------------------------------------------
\paragraph*{Notation.--} We use the metric convention $(-,+,+,+)$ in $D=4$ spacetime dimensions. Curvature-related quantities will be defined following Misner-Thorne-Wheeler's convention \cite{Wheeler1973}. The symbol $\eta_{ab}$ stands for the Minkowski metric, but written in an arbitrary coordinate system. $\nabla$ is the covariant derivative associated with the flat metric, and $\text{d}\mathscr{V}_\eta$ is the volume element in Minkowski spacetime. In this paper, the term cosmological constant will always refer to the corresponding quantity in the Einstein field equations, and not to any parameter occurring in the gravitational action. The reason for this remark is that, while in general relativity these two notions coincide, this is no longer true in unimodular gravity or Weyl-transverse gravity.
%------------------------------------------------

%------------------------------------------------
\section{The cosmological constant problem \label{sec:ccsec}}
%------------------------------------------------

The combination of general relativity and the standard model makes sense as an effective field theory up to, in principle, the Planck energy scale \cite{Donoghue1994,Burgess2004}. This framework even leads to specific predictions concerning genuine quantum corrections on different physical processes \cite{Bjerrum-Bohr2013,Donoghue2015,Donoghue2015b,Bjerrum-Bohr2016}, though none of these have been verified up to date due to their smallness. However, this effective field theory displays a famous feature: the so-called cosmological constant problem. That this is a recurrent problem in contemporary theoretical physics is demonstrated by the number of available reviews about it; see \cite{Weinberg1989,Burgess2013,Polchinski2006,Martin2012} for a small sample. It is not our aim to study all the different aspects, suggestions and ramifications of this problem. On the contrary, we will give a precise (and simple) mathematical meaning to the problem and keep our following discussion within this framework.

The root of the problem lies in the gravitating properties of the quantum vacuum. Let us first discuss the concept of the quantum vacuum in the absence of gravity, that is, in the framework of flat-spacetime quantum field theory. The quantum vacuum corresponds to the Poincar{\'e}-invariant (whether or not this symmetry is emergent is irrelevant) state of lowest energy. On top of this state, one can define non-vacuum states with a definite number of particles by using the corresponding creation operators, and evaluate the transition amplitudes between different states in a perturbative fashion. These calculations are pictorially represented by Feynman diagrams. From all these perturbative processes, in this paper we are interested in those that preserve the vacuum state or, in other words, that do not contain physical particles. These correspond in terms of Feynman diagrams to vacuum bubbles: diagrams with no external legs that represent the (perturbative) view of the quantum vacuum as a `sea' of virtual particles \cite{Milonni1994}. 

In flat-spacetime quantum field theory, the linked-cluster theorem (see, for instance, \cite{Kopietz2010}) permits to show explicitly that vacuum bubbles do not contribute to correlation functions so that they do not have any physical consequence, and therefore lack any operational meaning. This changes drastically if we include gravity in the discussion by means of general relativity and consider the resulting effective field theory. The decoupling of vacuum bubbles of the matter sector no longer holds as a result of the dependence of the spacetime volume form on the gravitational field: diffeomorphism invariance implies the coupling of gravity to these diagrams. The subsequent effect can be explicitly shown to lead to the renormalization of the cosmological constant \cite{Martin2012}.

A nontrivial running of the cosmological constant would not directly be worrisome. There are other well-known quantities in physics that are renormalized, such as the electron charge in quantum electrodynamics, for instance. Indeed, in an effective field theory framework, for any coupling constant there may be a non-trivial renormalization group equation that links the value of the constant with the energy scale at which it is measured. The trouble comes from the specific form of the renormalization group equation that applies to the cosmological constant.

In general relativity, the cosmological constant renormalization group equation can be evaluated by a number of techniques, all of them giving equivalent results. For instance, one can evaluate the effective action of matter fields, with the introduction of a regulator $\mu$. Using the heat kernel expansion \cite{Vassilevich2003} one can easily take into account the necessary counterterms that have to be added to renormalize the effective action. If $\Lambda_0$ is the bare cosmological constant, one gets then the equation \cite{Visser2002}
\begin{equation}
\Lambda=\Lambda_0+C_1\ln\left(\frac{\mu^2}{C_2}\right).\label{eq:ccreneq}
\end{equation}
The occurrence of logarithms is due to the use of dimensional regularization to regulate the divergent integrals; using a hard cutoff would imply the presence of powers of the cutoff $\mu$ \cite{Visser2002}. In Eq. \eqref{eq:ccreneq} the values of $C_1$ and $C_2$ depend on the masses of the matter sector. For a simplified matter sector with only one massive particle with mass $m$ one has, up to irrelevant numerical factors, $C_1\sim G c\,m^4/\hbar^3$ and $C_2\sim m^2$. For more involved particle spectrums one gets several terms as the second one on the right-hand side of Eq. \eqref{eq:ccreneq}, with a sum for all the different particles.

Eq. \eqref{eq:ccreneq} sets the stage for the cosmological constant problem. First of all, the measured value for the cosmological constant is several orders of magnitude smaller than $G c\,m^4/\hbar^3$ for any of the particles of the standard model \cite{Burgess2013}. Therefore, a change of order of magnitude on the regulator $\mu$ leads to a very large running when compared to this observational value. As a consequence, if one believes Eq. \eqref{eq:ccreneq}, it seems difficult to justify the value that is measured in cosmology for the cosmological constant, at least without invoking severe fine tunings. Most importantly, it is not even necessary to go to cosmological observations to detect this problem. Even if this is not always stressed, an effective description of the physics in the solar systems demonstrates its existence \cite{Martin2012,Kagramanova2006}: experiments in the solar system constrain the possible values of the cosmological constant to be much smaller than the natural order of magnitude obtained from Eq. \eqref{eq:ccreneq} when the particular constants for the particle content of the standard model are used \cite{Martin2012}. We would like to stress that this tension should not be confused with the additional issue of explaining from first principles the value of the cosmological constant that is observed. Additional clarifications in this regard are given in Sec. \ref{sec:volov}.

An additional complication arises when the effect of phase transitions on the cosmological constant is taken into account. The standard formalism to describe phase transitions in the cosmological evolution of our universe leads to large shifts of the cosmological constant across these transitions. This is however a phenomenon of different nature than the renormalization group equation of the cosmological constant, therefore requiring a separate study which is out of the scope of this paper. Note also that, as Weinberg stresses \cite{Weinberg1989}, there is no observational evidence that refutes the (calculable) effects of phase transitions on the cosmological evolution of our universe through the corresponding changes on the effective cosmological constant. In other words, there is no evidence that it could not be much bigger in the past; one could say even the contrary, i.e., that this could conform with the nowadays standard inflationary picture. But even solar system observations lead to strong tensions with the renormalization group equation \eqref{eq:ccreneq}, as stressed above. Indeed, what prevents to accept that the cosmological constant is a parameter that has to be fixed by observations, as any other fundamental constant in physics as the electron charge or the gravitational constant, is this very same equation (and similar equations that are obtained for higher orders in perturbation theory). This explains our focus on this feature, a perspective that is shared by many reviews; see, for instance, \cite{Burgess2013,Martin2012}. 

%------------------------------------------------
\section{Self-interaction of a helicity $\pm2$ graviton}
%------------------------------------------------

In flat-spacetime quantum field theory, fundamental interactions are classified in terms of the unitary representations of the Poincar\'e group \cite{Wigner1939}. Within this classification, gravity is associated with the massless spin-2 representation. That the corresponding representation is massless implies that the only physical states of the particle that carries the gravitational interaction (the graviton) are those with helicity $\pm2$. In field-theoretical terms, we can embed this representation into a second-rank symmetric tensor field $h^{ab}$ satisfying the equations of motion \cite{Ortin2004}
\begin{equation}
\square h^{ab}=0,
\end{equation}
as well as the constraints
\begin{equation}
\eta_{ab}h^{ab}=0,\qquad \nabla_b h^{ab}=0.\label{eq:cons}
\end{equation}
In fact, the physical objects are equivalence classes of $h^{ab}$ defined by the equivalence relation
\begin{equation}
{h'}^{ab}\sim h^{ab},\qquad {h'}^{ab}=h^{ab}+\eta^{ac}\nabla_c\xi^{b}+\eta^{bc}\nabla_c\xi^{a},\label{eq:fpsymm}
\end{equation}
where the generators $\xi^a$ satisfy
\begin{equation}
\nabla_a\xi^a=0,\qquad \square\xi^a=0.\label{eq:gencons}
\end{equation}
One can see that there always exists a generator $\xi^a$ such that the states with helicities $\sigma=0,\pm1$ are gauged away (for example, in the light-cone gauge \cite{Zwiebach2009}). On the other hand, the constraints \eqref{eq:cons} in the definition of the field $h^{ab}$ can be thought as the elimination of the scalar and vector representations of the Poincar\'e group (see Appendix I in \cite{Ogievetsky1965}). 

This on-shell description follows from the strict translation into field-theoretical terms of the requirement that gravity is associated with a helicity $\pm2$ graviton only, with no spin 1 or 0 content. One can take this description as the basis to discuss the self-coupling problem \cite{Barcelo2014}, although this procedure has some (tractable) drawbacks which can be avoided by slightly modifying the starting point. One can however relax the conditions \eqref{eq:cons} and \eqref{eq:gencons}, thus enlarging the gauge symmetries of the theory, which leads to Fierz-Pauli theory. This theory has Eq. \eqref{eq:fpsymm} as internal gauge invariance, now with no restrictions on the generators. However, it is remarkable that there exists an alternative extension which also reduces on-shell to a helicity $\pm2$ graviton \cite{Izawa1994}. The internal gauge symmetry of this alternative theory is
\begin{equation}
{h'}^{ab}\sim h^{ab},\qquad {h'}^{ab}=h^{ab}+\eta^{ac}\nabla_c\xi^{b}+\eta^{bc}\nabla_c\xi^{a}+\phi\eta^{ab},\label{eq:wtsymm}
\end{equation}
with generators satisfying only the first condition in \eqref{eq:gencons}, that is $\nabla_a\xi^a=0$, and $\phi$ is an arbitrary scalar function. The action of this theory is given by:
\begin{eqnarray}
\mathscr{A}_{\rm W}=\frac{1}{4}\int\text{d}\mathscr{V}_\eta\,\Big[2\eta_{bj}\delta^a_k\delta^i_c-\eta^{ai}\eta_{bj}\eta_{ck}
%\nonumber\\
-\eta_{bc}\delta^a_k\delta^i_j+\frac{3}{8}\eta^{ai}\eta_{bc}\eta_{jk}\Big]\nabla_ah^{bc}\nabla_ih^{jk}.\label{eq:freewtdiff}
\end{eqnarray}
This field-theoretical description and Fierz-Pauli theory are the only two linear extensions compatible with the on-shell description of gravitons given above \cite{Izawa1994,Alvarez2006}. This fact alone is interesting enough to explore this last theory to its ultimate consequences. Even if both are by construction completely equivalent as linear theories, their nonlinear completions may differ substantially. Note that the second kind of transformations in Eq. \eqref{eq:wtsymm} correspond to linearized Weyl transformations. Hence the action \eqref{eq:freewtdiff} represents the starting (i.e., linear) point for a Weyl-invariant description of gravity.

%------------------------------------------------------------------
\subsection{Previous results on the self-coupling problem}
%------------------------------------------------------------------

Let us discuss briefly some previous results concerning the recovery of general relativity from the picture sketched above. In order to describe how gravity affects matter, these linear theories should be coupled to matter fields. Let us focus for the moment on Fierz-Pauli theory. The relevant quantity to which these linear theories must couple is the stress-energy tensor of both matter fields and gravity, which implies the nonlinear nature of the ultimate theory \cite{Ortin2004}. The action that arises from this self-coupling procedure can be written as
\begin{equation}
\mathscr{A}=\mathscr{A}_2+\mathscr{A}_{\text{I}}.
\end{equation}
In this equation, $\mathscr{A}_2$ is the quadratic part that of the linear theory discussed above, while $\mathscr{A}_{\text{I}}$ is the nonlinear part that describes the self-interactions of gravitons. From this complete action, the stress-energy tensor could be obtained (using Hilbert's prescription). The nonlinear part of the action $\mathscr{A}_{\text{I}}$ is, by definition, fixed by requiring that it leads to a source in the equations of motion that is precisely this stress-energy tensor. In terms of equations, this requirement is translated into
\begin{equation}
\frac{\delta\mathscr{A}_{\text{I}}}{\delta h^{ab}}=\lambda\lim_{\gamma\rightarrow\eta}\frac{\delta(\mathscr{A}_2+\mathscr{A}_{\text{I}})}{\delta\gamma^{ab}}.\label{eq:app1eq1}
\end{equation}
Here $\lambda$, is the self-coupling constant. In order to obtain the iterative equations for the self-coupling problem \cite{Gupta1954}, we just have to expand $\mathscr{A}_{\text{I}}=\sum_{n=3}^\infty\lambda^n\mathscr{A}_n$. Comparing different orders in $\lambda$, one arrives to
\begin{equation}
\frac{\delta\mathscr{A}_{n+1}}{\delta h^{ab}}=\lim_{\gamma\rightarrow\eta}\frac{\delta\mathscr{A}_{n}}{\delta\gamma^{ab}},\qquad n\geq2.\label{eq:cons3}
\end{equation}
The partial actions on the right-hand side of these equations are written in terms of an auxiliary metric $\gamma_{ab}$. This procedure is not unique, as different choices of non-minimal couplings can be made; the importance of this issue for the self-coupling problem was discussed in \cite{Barcelo2014}. Using this formalism, it is possible to show explicitly that general relativity satisfies these equations (see the discussion just below). On the other hand, it is well-known that general relativity leads to Fierz-Pauli theory at the lowest order. Different non-minimal coupling prescriptions lead however to an entire one-parameter family of solutions. General relativity theory is then the only member in this family which preserves the number of generators of gauge symmetries that were present in the linear description of the gravitational field.

Integrating the iterative equations permits to write down explicitly their one-parameter family of solutions. However, once these expressions are known it is possible to show in a simpler way that these are solutions of the iterative equations. Making an analogy with differential equations, it is much easier to show that a function is a solution of a given differential equation, than integrating the latter to obtain its general solution. The necessary formalism was developed for the self-coupling problem in \cite{Butcher2009}, in which the details are thoroughly explained. This formalism introduces a functional $F$ on a tensor field $\gamma^{ab}+\lambda h^{ab}$, with $\gamma_{ab}$ an auxiliary metric. The resulting expression is expanded as a Taylor series on the deviations $\lambda h^{ab}$ from $\gamma^{ab}$, and then evaluated in $\gamma^{ab}=\eta^{ab}$. One can show then that the action is by construction a solution of the iterative equations. The relevant structure behind this proof can be highlighted using a single-variable function $F(\gamma+\lambda h)$ (tensor indices can be explicitly considered below but they do not introduce anything else relevant). The Taylor series of this function is
\begin{equation}
F(\gamma+\lambda h)=\sum_{m=0}^\infty \lambda^mF_m=\sum_{m=0}^\infty\frac{1}{m!} \frac{\partial^m F(\gamma)}{\partial \gamma^m}(\lambda h)^m.
\end{equation}
The elements of the set $\{F_m\}_{m=0}^\infty$ verify the relations
\begin{equation}
\frac{\partial F_{m+1}}{\partial h}=\frac{1}{m!}\frac{\partial^{m+1}F}{\partial \gamma^{m+1}}h^{m}=\frac{\partial F_{m}}{\partial\gamma},
\end{equation}
which are reminiscent of the iterative equations \eqref{eq:cons3}. The same happens for the ``action''
\begin{equation}
A=\int\text{d}^4x\,F(\gamma+\lambda h)(\nabla' h)^2=\sum_{m=0}^\infty\lambda^m\int\text{d}^4x\,F_m(\nabla' h)^2=\sum_{m=0}^\infty \lambda^mA_m.
\end{equation}
Note that the determinant of the field $\gamma^{ab}+\lambda h^{ab}$ that is necessary in order to properly define the integration in the equation above would be included in the term $F(\gamma+\lambda h)$. Also $\nabla'h$ should be symbolically understood as the covariant derivative with respect to $\gamma$. Then, discarding some irrelevant boundary terms one has the following equations that are formally equivalent to the iterative equations \eqref{eq:cons3}:
\begin{equation}
\frac{\partial A_{m+1}}{\partial h}=\int\text{d}^4x\,\frac{\partial F_{m+1}}{\partial h}(\nabla' h)^2=\int\text{d}^4x\,\frac{\partial F_{m}}{\partial\gamma}(\nabla' h)^2=\frac{\partial A_{m}}{\partial\gamma}.
\end{equation}
Boundary terms come from variations of the term $(\nabla' h)^2$ with respect to $h$ and $\gamma$. For the complete proof in the case of general relativity, following tightly these steps above, we refer the reader to \cite{Butcher2009}.

%------------------------------------------------------------------
\subsection{Self-coupling of the Weyl-invariant theory \label{sec:selfcweyl}}
%------------------------------------------------------------------

In this section we describe the nonlinear theory known as Weyl-transverse gravity and show explicitly that it satisfies the iterative equations \eqref{eq:cons3}, while it reduces to the explicitly Weyl-invariant linear description of gravitons with action \eqref{eq:freewtdiff} at the lowest order in $\lambda$. As sketched in the previous section, from previous analysis of the self-coupling problem starting from the linear Fierz-Pauli theory one expects that adding surface terms to the action \eqref{eq:freewtdiff} will be of relevance to the self-coupling problem, leading again to an entire family of solutions when the right non-minimal couplings are introduced. Nonetheless, from all these solutions only one would preserve the same degrees of freedom or, equivalently, the same number of generators of gauge symmetries as displayed by the linear theory. While knowing the explicit form of all these solutions would be interesting, in this paper we are mainly interested in this particular solution. It is then enough for our purposes to construct the action of a nonlinear theory satisfying this requeriment using symmetry arguments, showing then that it solves indeed the iterative equations \eqref{eq:cons3}.

Let us therefore start with the action of unimodular gravity,
\begin{equation}
\mathscr{A}=\frac{1}{\lambda^2}\int\text{d}\mathscr{V}_\eta\,\mathscr{R}(\hat{g}).
\end{equation}
In this equation, $\mathscr{R}(\hat{g})$ is the Ricci scalar of a metric $\hat{g}_{ab}$ with fixed determinant $\mbox{det}(\hat{g})=\mbox{det}(\eta)$. We can understand $\hat{g}_{ab}$ as a tensor field that lives in a flat background. This can be made explicit exploiting the well-known fact \cite{Rosen1940} that the Ricci scalar can be expressed in terms of the covariant derivatives associated with the flat metric $\eta_{ab}$. Integrating by parts, it follows that the equation above can be equivalently written as
\begin{equation}
\mathscr{A}=\frac{1}{4\lambda^2}\int\text{d}\mathscr{V}_\eta\left(2\hat{g}_{bs}\delta^a_t\delta^r_c-{\hat{g}}^{ar}\hat{g}_{bs}\hat{g}_{ct}\right)\nabla_a{\hat{g}}^{bc}\nabla_r{\hat{g}}^{st}.
\end{equation}
The notation we are using, explicitly covariant, makes manifest the invariance under general coordinate transformations of this theory. On the other hand, this action is invariant by construction under transverse diffeomorphisms which, at the infinitesimal level, can be written as
\begin{equation}
\delta_\xi {\hat{g}}^{ab}=\mathcal{L}_\xi{\hat{g}}^{ab},\qquad \nabla_a\xi^a=0.\label{eq:diff1}
\end{equation}
These are the nonlinear version of the first part of the linear symmetry \eqref{eq:wtsymm} with transverse generators. The natural nonlinear deformation of the linear Weyl invariance in the linear theory are Weyl transformations of the spacetime metric. To include these let us define
\begin{equation}
\hat{g}_{ab}=\kappa^{-1/4}g_{ab},\qquad\qquad \kappa=\mbox{det}(g)/\mbox{det}(\eta).
\end{equation}
This imposition guarantees that the action is invariant under conformal transformations, infinitesimally given by
\begin{equation}
\delta g_{ab}=\delta\omega\, g_{ab}.\label{eq:conf1}
\end{equation}
We can equivalently write the action as a functional of $g_{ab}$, namely
\begin{align}
\mathscr{A}&=\frac{1}{4\lambda^2}\int\text{d}\mathscr{V}_\eta\,\kappa^{1/4}\Big[(2g_{bs}\delta^a_t\delta^r_c-g^{ar}g_{bs}g_{ct})\nabla_a g^{bc}\nabla_r g^{st}\nonumber\\
&+\frac{1}{\kappa}\delta^a_b\delta^r_c\nabla_ag^{bc}\nabla_r\kappa -\frac{1}{2\kappa}g^{ar}g_{bc}\nabla_ag^{bc}\nabla_r \kappa-\frac{1}{8\kappa^2}g^{ar}\nabla_a \kappa\nabla_r \kappa\Big].\label{eq:finaction}
\end{align}
Using this expression, it is possible to show explicitly that the transformations \eqref{eq:diff1} and \eqref{eq:conf1} are combined in a way that ensures that this action is invariant under transverse diffeomorphisms on $g_{ab}$ and, independently, under conformal transformations.

Let us now show that this nonlinear theory can be obtained through the self-coupling of gravitons initially described by the action \eqref{eq:freewtdiff}. To do that, we will extend the formalism in \cite{Butcher2009}, which was useful to prove the analogue result in the case of general relativity, to consider actions of the type:
\begin{align}
\mathscr{A}&=\frac{1}{4\lambda^2}\int\text{d}\mathscr{V}_\eta\,\left[M^{ar}_{\ \ bcst}(g,\eta)\nabla_a g^{bc}\nabla_r g^{st}+N^{ar}_{\ \ bc}(g,\eta)\nabla_a g^{bc}\nabla_r \kappa\right.\nonumber\\
&\left.+O(g,\eta) g^{ar}\nabla_a \kappa\nabla_r \kappa\right].\label{eq:genself}
\end{align}
These actions display an additional dependence on the flat metric $\eta_{ab}$. The meaning of this feature is the following: starting with a linear theory in a flat spacetime, it is in principle not mandatory from a purely logical perspective that the resulting nonlinear theory has forgotten this flat spacetime structure. For general relativity, the form of the action ultimately implies that the reference to a flat metric can be absorbed without phenomenological consequences. As we discuss below this is not true in the case of Weyl-transverse gravity, which will be of capital importance for our discussion of the renormalization of the cosmological constant and the associated problem.

Following the steps sketched in the previous section, let us introduce an auxiliary metric field $\gamma_{ab}$ through $g^{ab}=\gamma^{ab}+\lambda h^{ab}$ and perform an expansion in the parameter $\lambda$:
\begin{equation}
\mathscr{A}=\sum_{m=0}^\infty\lambda^m\mathscr{A}_m,\qquad \mathscr{A}_m=\frac{1}{m!}\left.\partial_\lambda^m\mathscr{A}\right|_{\lambda=0},
\end{equation}
where the derivative $\partial_\lambda$ is defined as
\begin{equation}
\partial_\lambda\mathscr{A}=\int\text{d}^4x\,h^{ab}\frac{\delta\mathscr{A}}{\delta \gamma^{ab}}.
\end{equation}
Successive application of the derivative $\partial_\lambda$ permits to obtain
\begin{equation}
\mathscr{A}_{m+1}=\frac{1}{m+1}\int\text{d}^4x\,h^{ab}\left.\frac{\delta\mathscr{A}_{m}}{\delta \gamma^{ab}}\right|_{\lambda=0}=\frac{1}{(m+1)!}\left.\left(\int\text{d}^4x\,h^{ab}\frac{\delta}{\delta \gamma^{ab}}\right)^{m+1}\mathscr{A}\right|_{\lambda=0}.
\end{equation}
Two main observations stem from these relations. The first one is that
\begin{equation}
\frac{\delta \mathscr{A}_{m+1}}{\delta h^{ab}}=\left.\frac{1}{m!}\frac{\delta}{\delta \gamma^{ab}}\left(\int\text{d}^4x\,h^{cd}\frac{\delta}{\delta \gamma^{cd}}\right)^{m}\mathscr{A}\right|_{\lambda=0}=\frac{\delta \mathscr{A}_{m}}{\delta \gamma^{ab}}.
\end{equation}
Therefore, in the $\gamma\rightarrow\eta$ limit one recovers the iterative equations \eqref{eq:cons3}. Most importantly, the entire set of partial actions can be of course obtained from the entire action $\mathscr{A}$, but also from the quadratic action $\mathscr{A}_2$:
\begin{equation}
\mathscr{A}_n=\left.\frac{2}{n!}\left(\int\text{d}^4x\,h^{ab}\frac{\delta}{\delta \gamma^{ab}}\right)^{n-2}\mathscr{A}_2\right|_{\lambda=0}.\label{eq:iterecur}
\end{equation}
Here we use the label $n\geq 2$ in order to stress that both $\mathscr{A}_0$ and $\mathscr{A}_1$ are irrelevant to the construction: $\mathscr{A}_0$ is independent of $h_{ab}$ while $\mathscr{A}_1$ is identically zero in the $\gamma\rightarrow\eta$ limit.

For the family of actions \eqref{eq:genself}, the quadratic action reads in the flat limit
\begin{align}
\lim_{\gamma\rightarrow\eta}\mathscr{A}_2&=\frac{1}{4}\int\text{d}\mathscr{V}_\eta\,\left[M^{ar}_{\ \ bcst}(\eta,\eta)\nabla_a h^{bc}\nabla_r h^{st}\right.\nonumber\\
&\left.-N^{ar}_{\ \ bc}(\eta,\eta)\eta_{st}\nabla_a h^{bc}\nabla_r h^{st}+O(\eta,\eta) \eta^{ar}\eta_{bc}\eta_{st}\nabla_a h^{bc}\nabla_r h^{st}\right].\label{eq:lead}
\end{align}
The expression of this first nontrivial order in terms of a general auxiliary metric $\gamma_{ab}$ gives the non-minimal couplings which are necessary to the consistency of the formalism and, thus, the source to which the field $h^{ab}$ couples at first order. Higher orders can be directly evaluated from this first order using Eq. \eqref{eq:iterecur} to construct the infinite series of partial actions $\{\mathscr{A}_n\}_{n=3}^{\infty}$ and the corresponding sources.

Now we are in position to come back to the action \eqref{eq:finaction}, which is a particular case of \eqref{eq:genself} with
\begin{align}
&M^{ar}_{\ \ bcst}(g,\eta)=2\kappa^{1/4}g_{bs}\delta^a_t\delta^r_c-\kappa^{1/4}g^{ar}g_{bs}g_{ct},\nonumber\\
&N^{ar}_{\ \ bc}(g,\eta)=\frac{1}{\kappa}\kappa^{1/4}\delta^a_b\delta^r_c-\frac{1}{2\kappa}\kappa^{1/4}g^{ar}g_{bc},\nonumber\\
&O(g,\eta)=-\frac{1}{8\kappa^2}\kappa^{1/4}.
\end{align}
Note that we are ignoring the symmetrization of these quantities, which is not essential due to the fact that in both Eqs. \eqref{eq:genself} and \eqref{eq:lead} these tensors are contracted with quantities that display the relevant symmetries. With these equations at hand, we can check that the leading order \eqref{eq:lead} is exactly the linear Weyl-invariant description of gravitons with action \eqref{eq:freewtdiff}. More explicitly,
\begin{equation}
M^{ar}_{\ \ bcst}(\eta,\eta)=2\eta_{bs}\delta^a_t\delta^r_c-\eta^{ar}\eta_{bs}\eta_{ct}
\end{equation}
gives the two first terms inside the brackets in Eq. \eqref{eq:freewtdiff}, while
\begin{equation}
N^{ar}_{\ \ bc}(\eta,\eta)=\delta^a_b\delta^r_c-\frac{1}{2}\eta^{ar}\eta_{bc}
\end{equation}
and
\begin{equation}
O(\eta,\eta)=-\frac{1}{8}
\end{equation}
lead respectively to the third and four terms. This finishes the proof that the linear action \eqref{eq:freewtdiff} generates through self-coupling the nonlinear action \eqref{eq:finaction} through self-coupling.

Let us now come back to the dependence of Eqs. \eqref{eq:finaction} and \eqref{eq:genself} on the Minkowski metric $\eta_{ab}$. The latter equation is generic enough to encompass arbitrary functional dependencies on $\eta_{ab}$. However, Eq. \eqref{eq:finaction} makes explicit that for Weyl-transverse gravity this additional dependence on the Minkowski metric $\eta_{ab}$ is only through its determinant. This just reflects that in order to define the action of Weyl-transverse gravity in a coordinate-invariant fashion one has to introduce an auxiliary non-dynamical volume element, which given the nature of the construction is instinctively identified with the Minkowski volume element. This auxiliary volume element remains inert while applying Hilbert's prescription to obtain the source of the nonlinear equations of motion at different orders. This makes these sources automatically traceless, in order to guarantee compatibility with Weyl invariance at every step (identified by a given power of the coupling constant $\lambda$) of the iterative procedure. This ensures the decoupling order by order of the contributions of the quantum vacuum that would otherwise renormalize the cosmological constant, as we explain in Sec. \ref{sec:stable}.

For the discussion in the next section we need to include matter explicitly. Due to the absence of self-coupling, the integration of the iterative equations for the matter part of the action is straightforward. The result is, however, not unique. We can consider two additional principles in order to break this degeneracy. The first principle is the version of the equivalence principle shown by Weinberg to be a consequence of Poincar\'e invariance in \cite{Weinberg1964b}: the coupling constant $\lambda$ is the same for all fields (including $h_{ab}$). The second principle follows from the observation that conformal invariance of matter is not realized in the low-energy physics that we experience everyday. Then, it makes sense to ensure that matter fields are not affected by Weyl transformations \eqref{eq:conf1} (in other words, matter fields are inert under Weyl transformations). In practice, these principles permit to obtain the final matter action by replacing $\eta_{ab}$ with the composite field $\hat{g}_{ab}$. Note that both non-minimal couplings to $\hat{g}_{ab}$ and non-zero masses in the matter sector are allowed by construction.

%------------------------------------------------
\section{Keeping the cosmological constant small at all scales \label{sec:keepcc}}
%------------------------------------------------

The main conclusion that can be drawn from the previous discussion is that both general relativity and Weyl-transverse gravity are solutions to the self-coupling problem. That is, both nonlinear theories (i) describe gravitons in the non-interacting limit, and (ii) can be constructed explicitly by considering the self-interactions of this particle. Their only difference resides in their gauge groups. And this very feature explains the results of previous analyses about the uniqueness of general relativity as a solution to the self-coupling problem. Indeed, previous works always assumed that the characteristic gauge symmetry of gravitons is precisely that of Fierz-Pauli theory. For instance, in \cite{Boulware1975} the Ward identities associated to the gauge symmetries of Fierz-Pauli theory are an essential part of the demonstration of the uniqueness of general relativity. Our results are perfectly compatible with these in the sense that, if accepting this assumption about the gauge symmetries, general relativity arises as the only consistent nonlinear theory that preserves the original gauge invariance.  Nevertheless, we think it is important to stress that assuming that Fierz-Pauli theory is the correct description of linear gravitons is an additional assumption which, moreover, is in fact strongly suggested by the prior knowledge general relativity (and its gauge symmetries). This uncovers an additional non-uniqueness at the heart of the self-coupling problem of gravitons.

It is therefore worth exploring, from an emergent perspective, the possibility that the effective low-energy description of the gravitational interaction is given by Weyl-transverse gravity. The determination of the potential differences between these two choices represents an interesting field of study. It is convenient to review briefly the form of the classical field equations in Weyl-transverse gravity. Weyl symmetry can be exploited in order to fix a gauge in which the field equations take the same form as the traceless Einstein field equations in unimodular gravity: 
\begin{equation}
\mathscr{R}_{ab}-\frac{1}{4}\mathscr{R}g_{ab}=\bar{\kappa}\left(T_{ab}-\frac{1}{4}Tg_{ab}\right).
\end{equation}
These correspond to nine partial differential equations ($\bar{\kappa}=2\lambda^2$ where $\lambda$ is the coupling constant introduced previously). As explained in detail in \cite{Ellis2011,Ellis2013}, under the condition of the covariant conservation of the source $T_{ab}$ (that makes for a tenth equation) one recovers the full set of Einstein field equations, with $\Lambda=(\mathscr{R}+\bar{\kappa} T)/4$ an integration constant,
\begin{equation}
\mathscr{R}_{ab}-\frac{1}{2}\mathscr{R}g_{ab}+\Lambda g_{ab}=\bar{\kappa} T_{ab}.\label{eq:eequs}
\end{equation}
In the introduction we stressed that, in this paper, the cosmological constant is the parameter $\Lambda$ that appears in the equation just above. In general relativity, this quantity is directly linked to a coupling constant in the gravitational action but, in Weyl-transverse gravity, there is not such a connection. That is, these two parameters (the one in the action and the one in the field equations) are different.

Even taking into account the degeneracy on the phenomenology of these two theories at the classical level when certain conditions are met (namely conservation of the stress-energy tensor), there exists the possibility that differences are triggered by quantum effects. We may use the following clear analogy: radiative corrections can be understood as perturbations with respect to the tree-level physics; these differences would be equivalent to the (quite common) degeneracy breaking by perturbations in eigenvalue problems. In fact, this is exactly the situation in Weyl-transverse gravity \cite{Carballo-Rubio2015}: the cosmological constant is stable against radiative corrections in Weyl-transverse gravity, in stark difference with the situation in general relativity. This has been determinated evaluating the renormalization group of the coupling constants in the Weyl-transverse gravity action in the presence of quantum matter fields.

It has been recently argued \cite{Padilla2014} (see also \cite{Padilla2015}) that unimodular gravity cannot alleviate the cosmological constant problem. Given that this contradicts the main message of this paper (as well as previous works such as \cite{Ellis2011,Ellis2013}), it is convenient to dissect the arguments provided by these authors. The first comment in \cite{Padilla2014,Padilla2015} is that the unimodular condition must be implemented via a suitable Lagrange multiplier that becomes the effective cosmological constant, and that this renders its value radiatively unstable. First of all, using Lagrange multipliers to define unimodular gravity is not necessary nor convenient, as the theory can be defined simply over the space of unimodular metrics (this is, for instance, the approach in \cite{Weinberg1989}). The formulation used in the present paper provides even a more clear counterexample, as the action of Weyl-transverse gravity is defined over the space of conformal structures, which is directly defined on purely geometric grounds \cite{Ehlers2012} (and, obviously, without requiring the introduction of any sort of Lagrange multiplier). Moreover, we have shown that the Weyl-transverse formulation does not need to be postulated, but rather is derived unambiguously when completing the linear spin-2 representation of the Poincar\'e group. Together with its clean definition in terms of geometric notions, this makes this formulation of unimodular gravity the most fundamental one. Aside from the artificial nature of the representation that uses Lagrange multipliers, this representation does not realize at the off-shell level the characteristic shift symmetry of the field equations of unimodular gravity that is behind the decoupling of radiative corrections, as described in Sec. \ref{sec:stable} below. As quantum fluctuations are generally off-shell, different off-shell extensions of the same field equations may display different properties regarding radiative corrections. The second argument by these authors is rooted in a naive interpretation of unimodular gravity as a theory in which the determinant of the metric is fixed to be 1. Then, they argue, diffeomorphisms that are not volume-preserving would make a non-trivial Jacobian appear in the action, coming from the measure $\text{d}^4x$. However, the authors are missing the well-known result that unimodular gravity is in fact defined in a coordinate-invariant way in terms of a non-dynamical volume form \cite{Unruh1988} (which arises naturally in our previous discussion as $\text{d}\mathscr{V}_\eta$). This volume form `eats up' the Jacobian and remains, by construction, non-dynamical in all coordinate systems. Overall, the corollary of the discussion in this paragraph is that thinking about unimodular gravity as a theory in which the determinant of the metric is fixed to be 1 on-shell could be misleading, and that it is convenient to use instead its geometric definition following which the conformal (i.e., light-cone) structure of spacetime is dynamical, but the volume form is not.

Let us take a step back and consider a more general discussion concerning the minimal requirements that a theory has to verify in order to avoid the cosmological constant problem, and argue that Weyl-transverse gravity displays all of them. We shall also make some brief comments on a proposal by Volovik to fix the value of the cosmological constant that fits naturally in this framework and points to new avenues for future work.

%------------------------------------------------
\subsection{Stabilizing the cosmological constant \label{sec:stable}}
%------------------------------------------------

That Eq. \eqref{eq:ccreneq} does not make appearance in Weyl-transverse gravity can be seen from different, but complementary perspectives. As discussed in \cite{Carballo-Rubio2015}, an alternative way of understanding this feature is recalling the symmetries of Weyl-transverse gravity, in particular the Weyl symmetry. However, the action of Weyl-transverse gravity (plus matter) presents an additional global symmetry, the occurrence of which is intimately related to its local symmetries, that offers a complementary view on the relation between the interplay between the quantum vacuum and the gravitational interaction in this theory.

When considering a field theory on flat spacetime, there is a global symmetry that tell us that only relative energies have physical meaning, namely the shift symmetry
\begin{equation}
\mathscr{L}\longrightarrow \mathscr{L}+C_0,\label{eq:shift}
\end{equation}
where $C_0\in\mathbb{R}$ is a constant. One can trace back to this symmetry the decoupling of vacuum bubbles from correlation functions, and therefore from physical observables, that we have discussed in Sec. \ref{sec:ccsec}. The shift symmetry \eqref{eq:shift} is broken with the introduction of general relativity. Again, it is the consideration of a spacetime volume form that depends on physical fields the reason behind this feature. We have seen in Sec. \ref{sec:selfcweyl} that this feature arises naturally in the self-coupling problem when using a Weyl-invariant linear description of gravitons.

It is therefore clear that maintaining the shift symmetry \eqref{eq:shift} is a necessary condition to deal satisfactorily with the cosmological constant problem (see also \cite{Smolin2009,Padmanabhan2014b}). In order to guarantee this condition, there must exist a background volume form $\bm{\omega}$ so that the would-be cosmological constant term in the action is rendered innocuous to the classical dynamics.\footnote{It is also possible that $\bm{\omega}$ depends on the physical fields, but its integral is a topological invariant.} On the other hand, the classical field equations must contain a cosmological constant in order to match with cosmological observations. If we assume no deviations from classical physics, which is a well-motivated assumption given the quantum-mechanical nature of the cosmological constant problem, these equations should take the form \eqref{eq:eequs}. As explained above, this is indeed the case in Weyl-transverse gravity (taking into account the conservation of the stress-energy tensor of matter fields).

In terms of the parameters in the field equations \eqref{eq:eequs}, the fact that the shift transformation \eqref{eq:shift} is a global symmetry of the theory is expressed as
\begin{equation}
T_{ab}\longrightarrow T_{ab}+C_0 g_{ab},\qquad \qquad \Lambda\longrightarrow \Lambda+\bar{\kappa} C_0.\label{eq:shift2}
\end{equation}
Therefore, the combination of the two necessary conditions to deal with the cosmological constant problem, namely that the shift symmetry \eqref{eq:shift} holds and that there exists an effective cosmological constant (or, in other words, that one essentially recovers the Einstein field equations), point to the symmetry \eqref{eq:shift2} as a necessary condition in order to avoid the cosmological constant problem. Note that the transformation \eqref{eq:shift2} does \emph{not} correspond to a symmetry in general relativity, as in that case $\Lambda$ corresponds to a coupling constant that cannot be affected by symmetry transformations; in Weyl-transverse gravity $\Lambda$ is an integration constant that acts as a label for different solutions and can be therefore shifted by means of genuine symmetry transformations.

Weyl-transverse gravity presents the symmetry \eqref{eq:shift2}. However, this theory goes further, in the following sense. In effective field theory, ensuring that the symmetry \eqref{eq:shift2} is present would not be enough: one has to guarantee that this feature is preserved once radiative corrections are taken into account. In other words, it is necessary to impose a symmetry that forbids the term $\sqrt{|g|}\,\Lambda$ in the Lagrangian density, as this would spoil the shift symmetry \eqref{eq:shift}. There is a natural symmetry to consider in this regard, namely constant rescalings of the gravitational field,
\begin{equation}
g_{ab}\rightarrow \zeta^2g_{ab},\qquad\qquad \zeta\in\mathbb{R}.\label{eq:scale}
\end{equation}
One has to be careful about the interplay between this symmetry and the other gravitational symmetries. In particular, the symmetry under longitudinal diffeomorphisms in general relativity must be broken in order to guarantee that \eqref{eq:scale} is a symmetry (while still having second-order field equations). Incidentally, this is also what ultimately permits a non-anomalous implementation of Weyl invariance and hence the protection of the cosmological constant term against radiative corrections \cite{Carballo-Rubio2015}. But this implies that the number of degrees of freedom in the gravitational sector will be the same only if an additional symmetry is imposed; in particular, if the global symmetry \eqref{eq:scale} is extended to a local symmetry. Hence, we see that the discussion in this section represents an alternative route to motivate Weyl-transverse gravity, as the simplest theory that contains the minimal ingredients that permits handling both the classical and quantum aspects of the cosmological constant problem.

At the light of this discussion, it is interesting to consider an alternative proposal that has been recently presented \cite{Kaloper2014}. In this proposal, the modification of just the purely global properties of general relativity permits to overcome the cosmological constant problem without changing the local physics. As discussed by the authors, this is again due to the presence of the symmetry \eqref{eq:shift2}. In this case, this symmetry is trivially satisfied due to the constraint
\begin{equation}
\Lambda=\frac{\bar{\kappa}\int\text{d}^4x\sqrt{g}\,T}{4\int\text{d}^4x\sqrt{g}}.\label{eq:state}
\end{equation}
It is then clear to see that this proposal ``overkills'' the cosmological constant problem: recall that the necessary condition to deal with this problem is that the shift symmetry \eqref{eq:shift2} holds. This is a statement that concerns only the transformation properties of the effective cosmological constant. But Eq. \eqref{eq:state} presents a stronger statement, as it imposes a constraint on the effective cosmological constant that is not necessary in order to guarantee that the shift transformation \eqref{eq:shift2} is a symmetry. All the odd features of this proposal, such as the inclusion of global variables and the corresponding causality violations (note that to evaluate the effective cosmological constant \eqref{eq:state} in a cosmological context one would have to know in advance the entire evolution of the universe in all its points), can be traced back to the constraint \eqref{eq:state}. What we want to emphasize is that these are additional features that are not needed at all to deal with the cosmological constant problem, as shown by the discussion of this section, and the example of Weyl-transverse gravity.

%------------------------------------------------
\subsection{Setting its value at high energies \label{sec:volov}}
%------------------------------------------------

Once the cosmological constant (the integration constant that enters the gravitational field equations) is shown to be unaffected by radiative corrections, one may think about the kind of principles that could fix its value. In principle, a more fundamental theory or principle, going further than the low-energy effective field theory description (Weyl-transverse gravity plus matter fields), would be needed in order to obtain further insights about the nature of the cosmological constant and fix its actual value, which should be the one used at low energies in the low-energy effective framework. These considerations explain the title of this brief section: it is natural to expect that the value of the cosmological constant can be obtained in a suitable ultraviolet completion.

A natural possibility is that this principle is related to the actual vacuum of the high-energy theory. In this regard, Volovik's proposal \cite{Volovik2009} is one of the most satisfactory proposals from a physical perspective one can find in the literature. This proposal fits quite naturally in the general theme of this paper of emergence as it indeed arises from the study of quantum liquids. The discussion here follows the original formulation by Volovik \cite{Volovik2001,Volovik2004,Volovik2009}, with just a subtle (but important) deviation: in the original discussion one gets to the conclusion that the effective field theory must fail dramatically when evaluating some quantities such as the running of the cosmological constant. Taking Weyl-transverse gravity instead as the description of the gravitational interaction permits to overcome this conclusion, thus reconciling the principles of effective field theory and Volovik's arguments in a self-consistent combination. Indeed, one may argue that Weyl-transverse gravity provides the most natural realization of some of the ideas of this author. In our opinion, the following argument is worth further exploring due to its potential to justify the measured value for the cosmological constant.

The overall argument rests only in the assumption that our universe as a whole can be modeled as a quantum liquid, to which one can apply standard thermodynamical considerations \cite{Volovik2009}. The pressure in the vacuum state $p_0$ can be evaluated as the variation of the vacuum energy $\langle 0|\hat{H}|0\rangle$, where $|0\rangle$ is the ground state and $\hat{H}$ the Hamiltonian operator, with respect to the change of the volume of the system $V$. This leads to the equation
%F
\begin{align}
&p_0=-\frac{d\langle 0|\hat{H}|0\rangle}{dV}=-\frac{d[V \epsilon(N/V)]}{dV}\nonumber\\
&=-\epsilon(n)+n\frac{d\epsilon(n)}{d n}=-\frac{1}{V}\langle0|\hat{H}-\mu\hat{N}|0\rangle.
\end{align}
In this equation, $\mu=d\epsilon/d n$ is the chemical potential of the system, $\hat{N}$ the number operator with $N$ its mean value in the ground state, $\epsilon=\langle 0|\hat{H}|0\rangle/V$ the mean energy density and $n=N/V$ the mean density of particles. The vacuum pressure is then controlled by the grand-canonical Hamiltonian operator $\hat{H}-\mu \hat{N}$. This leads to the observation that the ground state presents the equation of state $\rho_0=-p_0/c^2$, typical of the cosmological constant, with
\begin{equation}
\Lambda=\bar{\kappa}\rho_0 c^2=\frac{\bar{\kappa}}{V}\langle0|\hat{H}-\mu\hat{N}|0\rangle.\label{eq:vacene}
\end{equation}
This result is universal: it does not depend on the details of the Hamiltonian operator (regardless of the system being relativistic or not), the statistics of the atoms in the fluid nor the corresponding low-energy effective field theory \cite{Volovik2009}. Note that only liquid states can exist as isolated systems, that is, with no external pressures being applied.

Then, in the absence of external forces and neglecting surface terms, the pressure $p_0$ must be identically zero. Hence, the (effective) cosmological constant \eqref{eq:vacene} must also be zero. Deviations from the perfect equilibrium would induce a nonzero cosmological constant that could be compared with the observed value in cosmological scenarios, though more precise models have to be constructed in order to make definite assertions in this regard.

This observation must be accompanied by a justification of the mismatch between the results one gets from Eqs. \eqref{eq:ccreneq} and \eqref{eq:vacene}, when assuming that the low-energy effective field theory contains gravity as described general relativity. Volovik's argument is that the evaluation of the properties of the vacuum (in particular, the cosmological constant) is not responsibility of the low-energy effective field theory. Only the knowledge of the underlying high-energy theory, which in this case is assumed to take the form of a condensed-matter-like model (the specific model is yet to be constructed), can determine the properties of the cosmological constant. This implies a complete failure of the principles of effective field theory in this construction.

However, if the low-energy description of the gravitational interaction is assumed to be given by Weyl-transverse gravity, both high- and low-energy perspectives on the cosmological constant smoothly match. The low-energy effective theory shows unequivocally that the cosmological constant is alien to the effective description, representing a fundamental parameter that is unaffected by radiative corrections. In other words, Weyl-transverse gravity is the natural realization of the observation that the low-energy effective field theory should be oblivious to the properties of the quantum vacuum, but for a set of physical constants that are non-calculable in the framework of the effective low-energy theory. The (effective) cosmological constant is one of these non-calculable physical constants. Therefore, its value has to be fixed in a wider framework, namely a suitable ultraviolet completion of Weyl-transverse gravity.

%------------------------------------------------
\section{Conclusions\label{sec:conc3}}
%------------------------------------------------

In this paper we have stressed that Weyl-transverse gravity is the first known example in the literature of what we can call ``minimal'' solution to the cosmological constant problem. The effective cosmological constant that appears in the field equations can take arbitrary, but radiatively stable values as it is protected by symmetries. Modifications with respect to the classical predictions of general relativity are only triggered by quantum effects, so that tree-level physics is preserved while one-loop and further corrections to the quantum vacuum are suppressed. Due to these properties, the criteria demanded in \cite{Burgess2013} for a satisfactory solution to the cosmological constant problem are verified. From the perspective of the low-energy physics, the cosmological constant is in this framework as mysterious as any other parameter in physics, such as the electron charge or the gravitational constant, but certainly not more mysterious than these. We have also emphasized that Weyl-transverse gravity is the most natural example illustrating previous arguments by Volovik concerning the nature of the cosmological constant and the quantum vacuum. These arguments are based in the potential emergence of the gravitational interaction from condensed-matter-like models that describe our universe as a quantum liquid, and point to the impossibility of describing the gravitational properties of the (true) quantum vacuum within the realm of the low-energy theory describing both gravity and matter fields. We have shown that this is precisely the situation in Weyl-transverse gravity. The structure of this theory may thus represent a first step in understanding the value of the cosmological constant from a set of fundamental principles, setting the basis for future work along these lines.

%------------------------------------
%Acknowledgements
%------------------------------------
 
\section*{Acknowledgements}

Financial support was provided by the Spanish MICINN through Projects No. FIS2014-54800-C2-1, FIS2017-86497-C2-1, FIS2014-54800-C2-2 and FIS2017-86497-C2-2 (with FEDER contribution), and by the Junta de Andaluc\'{i}a through Project No. FQM219. R.C.R. acknowledges support from CSIC through the JAE-predoc program, cofunded by FSE, and the Claude Leon Foundation in South Africa.

\bibliographystyle{elsarticle-num} 
\bibliography{absenceofccprob}	

%------------------------------------
\end{document}